# FIELD INDUCED SPIN REORIENTATION IN [Fe/Cr]$_n$ MULTILAYERS STUDIED BY NUCLEAR RESONANCE REFLECTIVITY

M. Andreeva[1*], A. Gupta[2], G. Sharma[3], S. Kamali[4], K. Okada[5] and Y. Yoda[5]

[1]*Faculty of Physics, M.V. Lomonosov Moscow State University, 119991, Russia;*
[2]*Indore Center, UGC-DAE Consortium for Scientific Research, Indore 452017, India;*
[3]*Center for Spintronic Materials, Amity University UP, Noida 201313, India;*
[4]*Department of Applied Science, University of California, Davis, California 95616, USA;*
[5]*Japan Synchrotron Radiation Research Institute, SPring-8, Sayo, Hyogo 679-5198, Japan*
[*]e-mail: Mandreeva1@yandex.ru

## Abstract

We present the depth-resolved nuclear resonance reflectivity (NRR) studies of the magnetization evolution in [$^{57}$Fe(3 nm)/Cr(1.2 nm)]$_{10}$ multilayer under the applied external field. The measurements have been performed at the station BL09XU of SPring-8 at different values of the external field (0 – 1500 Oe). We apply the joint fit of the delayed reflectivity curves and the time spectra of the nuclear resonance reflectivity measured at different grazing angles for enhancement of the depth resolution and reliability of the results. For the first time we show that the azimuth angle, which is used in all papers devoted to the magnetization profile determination, has more complicated physical sense due to the partially coherent averaging of the scattering amplitudes from magnetic lateral domains. We describe the way how to select the true azimuth angle from the determined "effective azimuth angle". Finally we obtain the noncollinear twisted magnetization depth-profiles where the spin flop state appears sequentially in different $^{57}$Fe layers at increasing applied field.

## Introduction

The interest to antiferromagnet/ferromagnet [F/AF] multilayers (ML) had been started long ago after the discovery of the GMR effect in [Fe/Cr] structures [1]. The immediately developed applications of this effect in microelectronics stimulated the extensive investigations of such systems resulting in the Nobel prize for Albert Fert and Peter Grünberg [2]. However, up to now some behavior features of F/AF multilayers at the action of the applied fields have not been inquired properly.

Ladder-step magnetization curve, observed in [3] for [Fe/Cr]$_n$ multilayer, gives the authors the idea that the transition from the antiferromagnetic interlayer alignment to the ferromagnetic one under the action of the external field in such systems takes place as a layer-by-layer change of the sign of the magnetization. Similar model of the magnetization reorientation had been presented in the earlier paper [4]. Resembling sequent overturn of magnetization of ferromagnetic layers separated by nonmagnetic one was observed in three layer system by X-ray resonant magnetic scattering (XRMS) [5]. Polarized neutron reflectivity (PNR) and off-specular scattering was measured at the Institute Laue Langevin on the reflectometer ADAM from the sample [Cr(0.9 nm)/$^{57}$Fe(6.7 nm)]$_{12}$ of (100) orientation in the external magnetic field of 19.5 mT applied along an in-plane easy axis (001) after a saturation in a field of 1 T. The data gave the experimental evidence of the nonuniform twisted canted state in the spin-flop phase [6]. The canting angles are maximal in the end layers and progressively relax towards the middle of the ML from both sides. The presence of magnetic off-specular scattering (vanishing at saturation) observed in this experiment meant that the layer magnetization was laterally not homogeneous, but rather decomposed into domains. A visualization of the field evolution of the [Fe(14 A)/Cr(11 A)]$_{20}$ magnetic structure grown with a (211) orientation was obtained by the least square fitting of the PNR data [7]. The authors concluded that the obtained picture quite well reproduced the theoretical predictions of [8-10], in particular it was shown that the spin-flopped region started from the top layer and was moved toward the center of the SL resulting in two antiphase domains. The similar complicated picture of the magnetization reorientation under the applied field was tested by PNR experiment in [11] for [Fe(4 nm)/Cr(1.1 nm)]$_{22}$ superlattice with cubic crystalline anisotropy. Measurements were performed under the applied fields 600, 200, 60 and 30 mT in order to probe the magnetisation depth-profile. The finding of the authors based on the data fit was that they got the proof of the predicted picture of the magnetization reorientation in decreasing fields

     The theories so far developed [8-14] present a rather complicated picture of the layer-by-layer reorientation in the antiferromagnetic MLs under the applied field, however in some papers the evolution of the layer magnetization vectors is treated with the assumption that magnetizations in all even and in all odd magnetic layers are

collinear in other words the reorientation of each magnetic sublattice takes place cophasingly. So the process of reorientation under the external field could be described just by two azimuth angles: each one for separate magnetic sublattice. In the paper [15] this model was used for the description of the hysteresis loops measured by the intensity variation of the half-order Bragg peak at the reflection of the soft $\pi$ linear polarized X-rays (at $L_3$ edge of Fe) from [Fe(1.52 nm)/Cr(2.56 nm)]$_{10}$ ML. In the paper [16] the variation of the layer magnetization directions in [Fe/FeO] superstructure was studied by nuclear resonance reflectivity (NRR) under gradual increase of the external field. The results were interpreted by an almost orthogonal moment alignment between adjacent Fe layers that are progressively rotated and finally collapsed to the direction of the applied filed. The result was later tested by PNR measurements [17].

In our paper we have performed the thorough investigation of the [$^{57}$Fe(3.0 nm)/Cr(1.2 nm)]$_{10}$. multilayer with antiferromagnetic interlayer coupling by means of the NRR using gradually increase of the external field. We have used the basic advantage of NRR experiments supplying the possibility to measure the time spectra of reflectivity at selected angles of incidence in addition to the NRR angular reflectivity curve. In such a way we got the combination of the spectroscopic and diffraction information in one experiment that supplies us with depth-profile information about hyperfine interactions in $^{57}$Fe layers and their magnetization direction. The joint fit of the angular curve and the time spectra of reflectivity, measured at several grazing angles, adds the reliability to the obtained magnetization depth-profiles.

We have made as well the essential correction to the previous ways of the data interpretation taking into account the partially coherent averaging of the reflected amplitudes from magnetic lateral domains. In this way we have been able to explain the results, obtained in two geometries: for transverse (T-geometry) and longitudinal (L-geometry) direction of the applied field relative SR beam.

**Theory**

Up to present the NRR experiments at synchrotron beamlines have been performed in the time domain: the short pulse of synchrotron radiation (SR) excites all hyperfine transitions in the resonant nuclei simultaneously and $\gamma$-quanta, attendant the decay of the excited states, are measured as a function of the delay time after prompt SR pulse (Nowadays with the developing of the nuclear monochromators it is possible to measure the spectra of reflectivity in the energy scale as common Mössbauer spectra [18-20]).

The time spectra of reflectivity $I(\theta,t)$ are calculated by applying the Fourier transform to the energy dependent $\sigma$- and $\pi$- reflectivity amplitudes.

$$I(\theta,t) = \left| \frac{1}{2\pi} \int_{-\infty}^{+\infty} R^{\pi \Rightarrow \sigma}(\theta,\omega) e^{-i\omega t} d\omega \right|^2 + \left| \frac{1}{2\pi} \int_{-\infty}^{+\infty} R^{\pi \Rightarrow \pi}(\theta,\omega) e^{-i\omega t} d\omega \right|^2. \quad (1)$$

Here we suppose that the incident SR field is $\sigma$-polarized, but the Mössbauer transition in $^{57}$Fe is of the magnetic dipole M1 type, so we follow the magnetic field of the radiation with the nuclei, which is $\pi$- polarized. Hyperfine interactions reveal themselves in the time spectra of reflectivity by the quantum beats well described for the nuclear resonance forward scattering, but the data interpretation is complicated by the dynamical beats and other effects specific for coherent decay of the excited nuclear system [21]. The time spectra of NRR are also distorted by the dynamical effects, but what is more essential by the phase relations between waves scattered by different layers. This last circumstance does not exist for the forward scattering and it provides us with the depth selective information relative the investigated parameters. The selectivity of the spectra measured at the different Bragg maxima to the depth distribution of hyperfine fields over one repetition period have been presented in [22] and thoroughly analyzed in [23]. In our case we are mostly interested in the magnetization profile in the whole ML.

The delayed nuclear reflectivity curve $I^{delayed}(\theta)$ is calculated by the integration of the time spectra over the whole delay time after the prompt SR pulse

$$I^{delayed}(\theta) = \int_{\Delta}^{T} I(\theta,t) dt \quad (2)$$

where $\Delta$ is the dead time of the APD detector and $T$ is the time interval between the successive SR pulses. Note that the actual value of $\Delta$ can essentially distort the behavior of $I^{delayed}(\theta)$ [24]. Normally the nuclear decay is virtually finished during the interval *T*, however if this interval is comparable with the lifetime of the excited nuclear state than "the tail" of the reflectivity decay from the previous pulse should be added to (2).

We used for the NRR data treatment our computer package REFTIM [25, 26]. Here we briefly present the used reflectivity theory. For the case of complicated noncollinear magnetic MLs we should work with the reflectivity matrices taking into account the possible polarization change of the radiation during multiple reflections-transitions through layer boundaries. Such matrix formalism has been developed in Optics [27, 28]. For plane waves $\sim \exp(i\frac{\omega}{c}\vec{\kappa}\vec{r} - i\omega t)$ we can describe the radiation field inside the multilayer in the following way:

$$\begin{pmatrix} \vec{E}(\vec{r},t) \\ \vec{H}(\vec{r},t) \end{pmatrix} = \begin{pmatrix} \vec{E}(z) \\ \vec{H}(z) \end{pmatrix} e^{i\left(\frac{\omega}{c}\vec{b}\vec{r} - \omega t\right)}, \qquad (3)$$

where $\vec{b}$ is the tangential component of the incident, reflected and transmitted wave vectors. We use the wave vectors $\vec{\kappa}$ in units of $\omega/c$, $\vec{\kappa} = \vec{b} + \eta\vec{q}$, where $\vec{q}$ is the unit vector normal to the surface (along $z$), so $|\vec{b}| = \cos\theta$, $\theta$ is the glancing angle for the incident radiation. Now we can work with the full radiation vectors $\vec{E}(z)$, $\vec{H}(z)$ not dividing them onto the eigen waves (4 in each of sublayers). It is known that it is worthy to go to the tangential components of $\vec{E}(z)$ and $\vec{H}(z)$, which are continuous at the transition through the boundary.

For the media characterized by the dielectric tensor $\hat{\varepsilon}(z) = 1 + \hat{\chi}(z)$, following [28], we present the Maxwell equations in a form:

$$\left(\vec{q}^{\times} \frac{d}{dz} + i\frac{\omega}{c}\vec{b}^{\times}\right)\begin{pmatrix} \vec{H}(z) \\ \vec{E}(z) \end{pmatrix} = i\frac{\omega}{c}\begin{pmatrix} -\hat{\varepsilon}(z)\vec{E}(z) \\ \vec{H}(z) \end{pmatrix} \qquad (4)$$

where superscript $^{\times}$ means the dual tensor, representing the vector product. Using the obvious scalar relations: $\vec{a}\vec{H} = \vec{q}\hat{\varepsilon}\vec{E},\ \vec{a}\vec{E} = -\vec{q}\vec{H}$, where $\vec{a} = \vec{b}\times\vec{q}$ we can get the relation between the total and tangential filed amplitudes determined by

$\vec{H}_t = \hat{I}\vec{H}$, $\vec{E}_t = \hat{I}\vec{E}$ ($\hat{I} = 1 - \vec{q} \circ \vec{q} = -\vec{q}^{\times 2}$ is the surface projective tensor, the sign $\circ$ means the outer product of vectors, subscript $_t$ means the tangential component):

$$\begin{pmatrix} \vec{H} \\ \vec{E} \end{pmatrix} = \begin{pmatrix} \hat{I} & -\vec{q} \circ \vec{a} \\ \dfrac{1}{\varepsilon_q} \vec{q} \circ \vec{a} & \hat{I} - \dfrac{1}{\varepsilon_q} \vec{q} \circ \vec{q} \hat{\varepsilon} \hat{I} \end{pmatrix} \begin{pmatrix} \vec{H}_t \\ \vec{E}_t \end{pmatrix}, \qquad (5)$$

where $\varepsilon_q = \vec{q}\hat{\varepsilon}\vec{q}$. Relation (5) allows us to diminish the number of variables in (4) from 6 to 4 and get finally the propagation matrix $\hat{M}$

$$\frac{d}{dz} \begin{pmatrix} \vec{H}_t(z) \\ \vec{q} \times \vec{E}(z) \end{pmatrix} = i\frac{\omega}{c} \hat{M}(z) \begin{pmatrix} \vec{H}_t(z) \\ \vec{q} \times \vec{E}(z) \end{pmatrix}, \qquad (6)$$

where

$$\hat{M} = \begin{pmatrix} \hat{A} & \hat{B} \\ \hat{C} & \hat{D} \end{pmatrix} \qquad (7)$$

and planar 2x2-blocks $\hat{A}, \hat{B}, \hat{C}, \hat{D}$ have the form ($\tilde{\tilde{\varepsilon}}$ is the transposed to the reciprocal tensor of $\hat{\varepsilon}$) [28]:

$$\hat{A} = \frac{1}{\varepsilon_q} \vec{q}^{\times} \hat{\varepsilon} \vec{q} \circ \vec{a}, \qquad \hat{B} = \frac{1}{\varepsilon_q} \hat{I}\tilde{\tilde{\varepsilon}}\hat{I} - \vec{b} \circ \vec{b},$$

$$\hat{C} = \hat{I} - \frac{1}{\varepsilon_q} \vec{a} \circ \vec{a}, \qquad \hat{D} = -\frac{1}{\varepsilon_q} \vec{a} \circ \vec{q}\hat{\varepsilon}\vec{q}^{\times}. \qquad (8)$$

For NRR we can simplify the general expressions (8), obtained in optics, if we take into account that $\hat{\chi} \sim 10^{-5}$ and reflectivity is essential only at grazing incidence angles $\theta < 10^{-2}$. We have got the very simple approximate expression for the differential propagation matrix applicable for NRR at grazing angles [29]:

$$\hat{M} \cong \begin{pmatrix} 0 & 0 & 1 & 0 \\ \vec{a}\hat{\chi}\vec{q} & 0 & 0 & \sin^2\theta + \vec{a}\hat{\chi}\vec{a} \\ \sin^2\theta + \vec{q}\hat{\chi}\vec{q} & 0 & 0 & \vec{q}\hat{\chi}\vec{a} \\ 0 & 1 & 0 & 0 \end{pmatrix}. \qquad (9)$$

The used coordinate system is presented in Fig. 1.

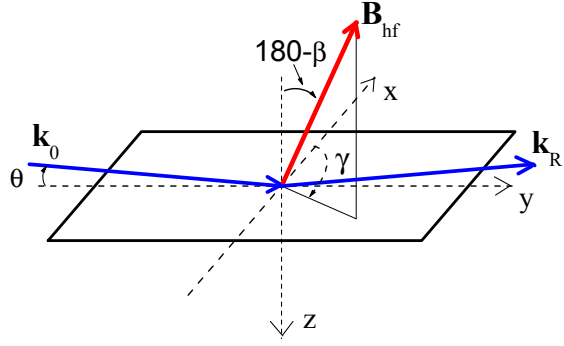

*Fig. 1. The used coordinate system.*

In some specific cases the simplified expression (9) is incorrect (some examples are considered in [30]) and the exact propagation matrix should be used.

The eigen values $\eta$ of the matrix (9), which at the same time are the normal components of the wave vectors in units of $\omega/c$, can be easily found from the biquadratic equation:

$$\eta^4 - (2\sin^2\theta + \vec{a}\hat{\chi}\vec{a} + \vec{q}\hat{\chi}\vec{q})\eta^2 + (\sin^2\theta + \vec{q}\hat{\chi}\vec{q})(\sin^2\theta + \vec{a}\hat{\chi}\vec{a}) - \vec{a}\hat{\chi}\vec{q}\,\vec{q}\hat{\chi}\vec{a} = 0, \quad (10)$$

so the integral of (6) in the layer of thickness $d$ with constant $\hat{\chi}(z) = const$ we calculate e.g. by the Sylvester formula [31]:

$$e^{i\frac{\omega}{c}d\hat{M}} = \sum_{j=1}^{4} e^{i\frac{\omega}{c}d\eta^{(j)}} \frac{\prod_{n \neq j}(\hat{M} - \eta^{(n)})}{\prod_{n \neq j}(\eta^{(j)} - \eta^{(n)})} \quad (11)$$

The analytical expression for the integral propagation matrix have also been found in [29] for such simple form of $\hat{M}$ (9) by interchanging of the lines and columns of $\hat{M}$:

$$\tilde{\tilde{L}}(d) = e^{ik\tilde{\tilde{M}}d} = \begin{pmatrix} \cos(kd\sqrt{\tilde{\tilde{B}}\tilde{\tilde{C}}}) & i\tilde{\tilde{C}}^{-1}\sqrt{\tilde{\tilde{C}}\tilde{\tilde{B}}}\sin(kd\sqrt{\tilde{\tilde{C}}\tilde{\tilde{B}}}) \\ i\tilde{\tilde{B}}^{-1}\sqrt{\tilde{\tilde{B}}\tilde{\tilde{C}}}\sin(kd\sqrt{\tilde{\tilde{B}}\tilde{\tilde{C}}}) & \cos(kd\sqrt{\tilde{\tilde{C}}\tilde{\tilde{B}}}) \end{pmatrix}, \quad (12)$$

were

$$\tilde{\tilde{B}} = \begin{pmatrix} \vec{a}\hat{\chi}\vec{q} & \sin^2\theta + \vec{a}\hat{\chi}\vec{a} \\ \sin^2\theta + \vec{q}\hat{\chi}\vec{q} & \vec{q}\hat{\chi}\vec{a} \end{pmatrix}, \quad \tilde{\tilde{C}} = \begin{pmatrix} 0 & 1 \\ 1 & 0 \end{pmatrix} \quad (13)$$

This algorithm is similar to the super-matrix method used in the PNR theory [32, 33], however the calculations of the 2x2 matrix functions in (12) (The detailed algorithm of these procedures with usage of the Pauli matrices has been presented e.g. in the early work devoted to the Faraday rotation for Mössbauer radiation [34]) take more computation time than the calculations by Silvester formula (11).

The known eigen values of $\hat{M}$ allow us to get as well the so called surface impedance planar tensors $\hat{\gamma}$, connected the tangential components of the radiation field $\vec{q} \times \vec{E}$ and $\vec{H}_t$, i.e. $\vec{q} \times \vec{E} = \hat{\gamma} \vec{H}_t$, which are needed for the boundary task. For the waves propagating in one direction (connected with the eigen values $\eta^{(1)}, \eta^{(2)}$):

$$\hat{\gamma}^M = \frac{1}{\sin^2\theta + \vec{a}\hat{\chi}\vec{a} + \eta^{(1)}\eta^{(2)}} \begin{pmatrix} \eta^{(1)}\eta^{(2)}(\eta^{(1)} + \eta^{(2)}) & \vec{q}\hat{\chi}\vec{a} \\ -\vec{a}\hat{\chi}\vec{q} & (\eta^{(1)} + \eta^{(2)}) \end{pmatrix} \quad (14)$$

The integral propagation 4x4-matrix for the whole multilayer $\hat{L}$ is calculated as the product of the propagation matrices of the individual layers $\hat{L}(d_n)$ because the tangential components of the radiation field are continuous at the layer boundaries. The reflectivity 2x2-matrix $\hat{r}$ for the tangential components of the magnetic filed of the radiation $\vec{H}_t^R = \hat{r} \vec{H}_t^0$ is calculated by the formula [29]:

$$\hat{r} = [\hat{\gamma}^D (\hat{L}_1 + \hat{L}_2 \hat{\gamma}^R) - (\hat{L}_3 + \hat{L}_4 \hat{\gamma}^R)]^{-1} [(\hat{L}_3 + \hat{L}_4 \hat{\gamma}^0) - \hat{\gamma}^D (\hat{L}_1 + \hat{L}_2 \hat{\gamma}^0)], \quad (15)$$

where $\hat{L}_{1,2,3,4}$ are the 2x2-blocks of $\hat{L}$ and superscripts $D,R$ refer to the substrate and the outward media. Taking into account that $\vec{E} = (\vec{q} \circ \vec{a} - \vec{q}^\times \hat{\gamma}) \vec{H}_t$ (from (5)), the components of the reflectivity matrix in $\sigma, \pi$ orts used in (1) are determined by the relations:

$$R^{\hat{\sigma} \Rightarrow \hat{\sigma}} = -r_{22}, \quad R^{\hat{\sigma} \Rightarrow \hat{\pi}} = -r_{12}\sin\theta, \quad R^{\hat{\pi} \Rightarrow \hat{\sigma}} = r_{21}/\sin\theta, \quad R^{\hat{\pi} \Rightarrow \hat{\pi}} = r_{11}. \quad (16)$$

For the angles much larger than the critical angle of the total external reflection the reflectivity can be described in the kinematical limit of the exact theory. For the scalar susceptibility of layers it has been accurately derived in [35] and for the case of anisotropic layers it is done in [36]:

$$\hat{R} \cong \sum_{j=1}^{L} e^{-i\frac{\omega}{c}d_1\hat{n}_1^-} ... e^{-i\frac{\omega}{c}d_{j-1}\hat{n}_{j-1}^-} \hat{r}_{j-1,j} e^{+i\frac{\omega}{c}d_{j-1}\hat{n}_{j-1}^+} ... e^{+i\frac{\omega}{c}d_1\hat{n}_1^+}, \tag{17}$$

For the refraction $\hat{n}_j^\pm$ and reflectivity $\hat{r}_{j-1,j}$ matrices the following expressions have been obtained:

$$\hat{r}_{j-1,j} = \frac{1}{4\sin^2\theta}(\hat{\chi}_{j-1}^\perp - \hat{\chi}_j^\perp), \qquad \hat{n}_j^\pm = \pm(\sin\theta + \frac{\hat{\chi}_j^{\pm\perp}}{2\sin\theta}). \tag{18}$$

For small grazing angles typically used in NRR experiments the transversal 2x2-matrices $\hat{\chi}_j^\perp$ in (18) can be calculated in the plane perpendicular to the beam direction (to vector $\vec{b}$, which is along the y-axis, see Fig. 1), then $\hat{\chi}_j^\perp$ can be presented in $\sigma, \pi$ orts as

$$\hat{\chi}_j^\perp = \begin{pmatrix} \vec{a}\hat{\chi}\vec{a} & -\vec{a}\hat{\chi}\vec{q} \\ -\vec{q}\hat{\chi}\vec{a} & \vec{q}\hat{\chi}\vec{q} \end{pmatrix} = \frac{\lambda^2}{\pi}\rho \hat{f}_{\vec{b}\to\vec{b}}, \tag{19}$$

Where we can put $|\vec{a}|=1$. In (20) we as well write down the known relation between the susceptibility tensor and the coherent forward scattering amplitude $\hat{f}_{\vec{b}\to\vec{b}}$, $\rho$ is the volume density of the scattering centers.

Unfortunately, in general case the matrix exponentials $e^{i\hat{n}kz}$ in (17) can not be rearranged, so this kinematical way of computation does not considerably speed-up the computations. Only if the polarization dependent absorption can be neglected, the expression (17) can be simplified to the expression (formally similar to that obtained in [35], but for matrices $\hat{\chi}_j^\perp$):

$$\hat{R} \cong \frac{1}{4\sin^2\theta} \sum_{j=1}^{L} e^{iQz_{j-1}}(\hat{\chi}_{j-1}^\perp - \hat{\chi}_j^\perp), \tag{20}$$

where $Q = \frac{4\pi}{\lambda}\sin\theta$ is the vector of scattering and $z_{j-1}$ is the z-coordinate of the $(j-1)/j$ boundary counting off the surface, $L$ is the number of the interlayer boundaries. Proceeding from this simplest expression (20) for the reflectivity amplitude, NRR experiments are often called as the forward scattering experiments.

However, the different phases ($e^{iQz_j-1}$ in the simplest case (20)) of the reflection amplitudes from layer boundaries lead to much more complicated shape of the reflectivity spectra than of the forward scattering spectra. But these phases supply us with the depth resolution. Kinematical approach is not applicable near the critical angle, however, the kinematical formula (19) gives us an opportunity for qualitative considerations of some effects.

In the matrix formalism of reflectivity the roughness of layer boundaries can not be included as the additional factor like it has been done for the scalar reflectivity theory (e.g. in [37]). For anisotropic layers we take into consideration the interface layers with variable profiles of the used parameters (electron density, density of nuclei characterizing by the definite kind of hyperfine interaction etc.). The smooth profiles are automatically partitioned in calculations by the determined number of steps. (The detailed description is given in [25]). The procedure of the roughness inclusion to the matrix reflectivity theory, described in [38], is much more complicated and uncertain for differential consideration of the separate hyperfine contributions.

The most lengthy description in calculations is needed for the susceptibility tensor $\hat{\chi}$ of the Mössbauer medium. It consists from two parts associated with the scattering by the electrons $\chi^{el}$ and resonant nuclei $\hat{\chi}^{nucl}$:

$$\hat{\chi} = \chi^{el} + \hat{\chi}^{nucl}. \qquad (21)$$

For the electronic part we have

$$\chi^{el} = -\frac{\lambda^2}{\pi} r_e \sum_n \rho_n (Z_n + \Delta f_n' + i\Delta f_n'') = -2\delta + i2\beta, \qquad (22)$$

where $r_e = 2.818 \cdot 10^{-6}$ nm is the classical electron radius, $\rho_n$ is the volume density of the n-type atoms, $Z_n$ is their atomic number, $\Delta f_n', \Delta f_n''$ are the anomalous dispersion corrections, $\delta$ and $\beta$ are the refraction and absorption terms in the expression of the refractive index $n = 1 + \delta + i\beta$. (We can find the Table values of $\delta$ and $\beta$ for some substances in e.g. [39]).

$$\hat{\chi}^{nucl}(\omega) = -\frac{\lambda}{2\pi}\frac{\Gamma_{nat}}{2}\sigma_{res}\frac{2L+1}{2I_e+1}\times$$

$$\times \sum_j \rho_j^{nucl} f_j^{LM} \sum_{m_e,m_g} \frac{\left|\langle I_g m_g L\Delta m | I_e m_e\rangle\right|^2}{\hbar\omega - E_{jR}(m_e,m_g) + \frac{i\Gamma}{2}} \vec{h}_{\Delta m}\cdot\vec{h}_{\Delta m}^* \quad (23)$$

where for $^{57}$Fe (M1 transition) $L=1$, $I_e=3/2$, $I_g=1/2$, $m_e, m_g$ are the magnetic quantum numbers, $\Delta m = m_e - m_g = \pm 1, 0$, $\langle I_g m_g L\Delta m | I_e m_e\rangle$ are the Clebsch–Gordan coefficients, $\sigma_{res}=2.56\ 10^{-4}\ \text{nm}^2$ is the resonance cross-section, $\lambda=0.086$ nm, $\Gamma_{nat}=0.097$ mm/s (4.665 neV), $j$ numerate the kind of the hyperfine splitting (multiplet number in Mössbauer spectrum), $f_j^{LM}$ is the Lamb–Mössbauer factor for this type of nuclei and $\rho_j^{nucl}$ is the volume density of the resonant nuclei possessing $j$ type of hyperfine splitting. $\hat{h}_{\Delta m}$ in (23) are the spherical orts of the hyperfine field principal axis:

$$\vec{h}_{\pm 1} = \mp i \frac{\vec{h}_x \pm i\vec{h}_y}{\sqrt{2}}, \quad \vec{h}_0 = i\vec{h}_z. \quad (24)$$

In the chosen coordinate system we have ($\beta$ and $\gamma$ are the polar and azimuth angles determining $\vec{B}_{hf}$ orientation as it shown in Fig. 1)

$$\vec{h}_x = (-\cos\beta\cos\gamma, -\cos\beta\sin\gamma, \sin\beta),$$
$$\vec{h}_y = (\sin\gamma, -\cos\gamma, 0),$$
$$\vec{h}_z = (\sin\beta\cos\gamma, \sin\beta\sin\gamma, \cos\beta). \quad (25)$$

The matrix (20) for different hyperfine transitions takes a form:

$$\hat{\chi}^{\perp}_{\Delta m=0} \propto \begin{pmatrix} \sin^2\beta\cos^2\gamma & -\sin\beta\cos\beta\cos\gamma \\ -\sin\beta\cos\beta\cos\gamma & \cos^2\beta \end{pmatrix} \quad (26)$$

$$\hat{\chi}^{\perp}_{\Delta m=\pm 1} \propto \frac{1}{2}\begin{pmatrix} \sin^2\gamma + \cos^2\gamma\cos^2\beta & (\cos\beta\cos\gamma \mp i\sin\gamma)\sin\beta \\ (\cos\beta\cos\gamma \pm i\sin\gamma)\sin\beta & \sin^2\beta \end{pmatrix} \quad (27)$$

It is easily seen that in the case of $\beta=0$ the transitions with $\Delta m = 0$ do not participate in the scattering the $\pi$-polarized radiation (that is the polarization of the magnetic field of the $\sigma$-polarized SR), the amplitude of scattering at the $\Delta m = \pm 1$ is not changed after substitution $\gamma$ by $(180-\gamma)$ (symmetrical relative the beam direction), the scattering to the changed polarization state is antisymmetric, but it is symmetric for scattering to the same polarization state relative substitution $\gamma$ by $(-\gamma)$ (in one-fold scattering approximation).

If the hyperfine splitting is absent $E_R(m_e, m_g) = E_R$, then

$$\frac{2L+1}{2I_e+1} \sum_{m_e,m_g} \left|\langle I_g m_g L \Delta m | I_e m_e \rangle\right|^2 \vec{h}_{\Delta m} \cdot \vec{h}^{*}_{\Delta m} = 1 \qquad (28)$$

and we come to the expression for the scalar $\chi^{nucl}$ given in e.g. [21]. It is useful to have the averaging of tensors $\vec{h}_{\Delta m} \cdot \vec{h}^{*}_{\Delta m}$ for the cases of the random orientation of the hyperfine fields in the surface plane ($\beta = 90^o$)

$$\overline{\vec{h}_{\pm} \cdot \vec{h}^{*}_{\pm}}^{\perp \vec{q}} = \frac{1}{4}(1 + \vec{q} \circ \vec{q}) = \frac{1}{4}\begin{pmatrix} 1 & 0 & 0 \\ 0 & 1 & 0 \\ 0 & 0 & 2 \end{pmatrix}, \qquad (29)$$

$$\overline{\vec{h}_{0} \cdot \vec{h}^{*}_{0}}^{\perp \vec{q}} = \frac{1}{2}(1 - \vec{q} \circ \vec{q}) = \frac{1}{2}\begin{pmatrix} 1 & 0 & 0 \\ 0 & 1 & 0 \\ 0 & 0 & 0 \end{pmatrix}; \qquad (30)$$

For the case of the random orientation of the hyperfine fields in 3D space (Averaging over sphere means $\frac{1}{4\pi}\int_0^{2\pi}\int_0^{\pi} ..... \sin\varphi \, d\gamma \, d\beta$) we have scalar nuclear susceptibility $\chi^{nucl}$:

$$\overline{\vec{h}_{\pm} \cdot \vec{h}^{*}_{\pm}} = \overline{\vec{h}_{0} \cdot \vec{h}^{*}_{0}} = \frac{1}{3}\begin{pmatrix} 1 & 0 & 0 \\ 0 & 1 & 0 \\ 0 & 0 & 1 \end{pmatrix}. \qquad (31)$$

The hyperfine splitting of the nuclear levels $E_R(m_e, m_g)$ is well known in Mössbauer spectroscopy. For the case of the uniaxial hyperfine fields we have.

$$E_R(m_e, m_g) = IS + B_{hf}(G^{ex} m_e - G^{gr} m_g) + \Delta_{EFG} * (A_Q^{ex} - A_Q^{gr} * C_Q), \quad (32)$$

where $IS$ (mm/s) is the isomeric shift, $B_{hf}$ (T) is the magnetic hyperfine field value, $G^{ex} = -0.067897$ (mm/s / T) and $G^{gr} = 0.118821$ (mm/s / T) are the gyromagnetic ratios for the excited and ground states, $\Delta_{EFG}$ (mm/s) is the electric quadrupole splitting ($\Delta_{EFG} = eQq/2$ in the notations of [40]), $A_Q^{ex,gr} = \dfrac{3m_{e,g}^2 - I_{e,g}(I_{e,g}+1)}{2I_{e,g}(2I_{e,g}-1)}$, $C_Q = \dfrac{Q^{gr}}{Q^{ex}}$, $Q^{ex} = 0.2$ barn and $Q^{gr} = 0$ are the quadrupole moments of the excited and ground states. The values are given for $^{57}$Fe. For the case of the noncollinear $B_{hf}$ and the principal axis of the electric field gradient or for the case of non-axially symmetric electric quadrupole interaction the eigen functions of the nuclear energy levels are the combination of the magnetic states and we follow the paper [41] for the description of $\hat{\chi}$ in REFTIM [25, 26].

Note that $\hat{\chi}^{nucl}(\omega)$ (as well as $\chi^{el}$ has depth profile) in multilayer is specific for each separate sublayer and this specification should be determined by the additional index number adding to $\rho_j^{nucl}$, to hyperfine axis orientation $\beta_j, \gamma_j$ and to possible portion of the random or plane orientation in each sublayer. So the description of the multilayer model includes the huge number of independent parameter for fit.

**Partial transverse coherence of the incident beam**

The most nontrivial description for the reflectivity is needed for the case when lateral inhomogendous magnetic structure takes place. Nowadays it is clear that even a very thin magnetic ML has lateral domains. A lot of experimental methods successfully visualize them. Up to now the most of the papers devoted to the NRR experiments (excluding the off-specular measurements, e.g. [42]) have presented the definite azimuth magnetization angle for each sublayer as the result of the reflectivity data fit [16, 43-45]. If the measurements are not performed in the saturation state of the sample that is in contradiction with the latest results on the domain structure visualizations for ultrathin multilayers (see e.g. [46-48]).

NRR measurements are executed at grazing angles $\sim 10^{-3}$ rad. Even the very narrow SR beam (~20 μm) illuminates a rather large area on the surface ~20000 μm (along the beam direction) which exceeds the typical lateral domain size. If in the absence of the external field we have the equal number of domains with different orientation of $\vec{B}_{hf}$ in the surface plane, we could suppose that we have the case (29)-(30), which for $\sigma$ polarized SR beam gives the identical results with the case $\beta=90º$, $\gamma=0º,180º$ [49-51]. But surprisingly the NRR data interpretation has given the definite azimuth angle for the magnetization direction which is difficult to explain especially for the polycrystalline film in e.g. [43].

Here we try to give the more realistic interpretation of the results. It takes into account the finite transverse coherence length of SR [52-53]. We should suppose that probably not all scattered amplitudes from different domains are added coherently in the reflectivity signal (Fig. 2).

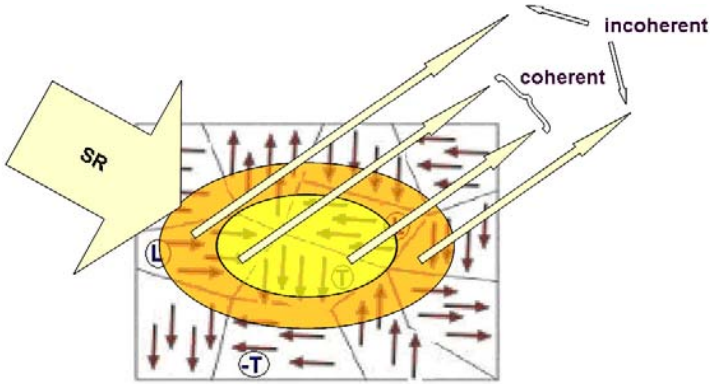

*Fig. 2. The illustration of the partially coherent interference of waves reflected by different domains.*

Assuming the 4-fold in-plane anisotropy we will operate with 4 possible $\sigma-$ and $\pi-$polarized reflected amplitudes $f_{\sigma,\pi}^{L,-L,T,-T}$ for domains with magnetization directions along the beam (L and -L) and perpendicular to the beam (T and –T). So the reflectivity intensity can be presented by the formula:

$$I_R(t) \propto (|\alpha f_\pi^L|^2 + |\beta f_\pi^{-L}|^2 + |\gamma f_\pi^T|^2 + |\delta f_\pi^{-T}|^2 +$$
$$+ 2C_{coh} \text{Re}\left(\alpha\beta f_\pi^L f_\pi^{-L*} + \alpha\gamma f_\pi^L f_\pi^{T*} + \alpha\delta f_\pi^L f_\pi^{-T*} + \beta\gamma f_\pi^{-L} f_\pi^{T*} + \beta\delta f_\pi^{-L} f_\pi^{-T*} + \gamma\delta f_\pi^T f_\pi^{-T*}\right))$$
$$+ (|\alpha f_\sigma^L|^2 + |\beta f_\sigma^{-L}|^2 + |\gamma f_\sigma^T|^2 + |\delta f_\sigma^{-T}|^2 +$$
$$+ 2C_{coh} \text{Re}\left(\alpha\beta f_\sigma^L f_\sigma^{-L*} + \alpha\gamma f_\sigma^L f_\sigma^{T*} + \alpha\delta f_\sigma^L f_\sigma^{-T*} + \beta\gamma f_\sigma^{-L} f_\sigma^{T*} + \beta\delta f_\sigma^{-L} f_\sigma^{-T*} + \gamma\delta f_\sigma^T f_\sigma^{-T*}\right))$$

(33)

where $\alpha, \beta, \gamma, \delta$ are the relative amount of the $L, -L, T, -T$ domains respectively. In (33) we insert the factor $C_{coh}$ ($0 \leq C_{coh} \leq 1$) which supplies the suppression of the interference of the waves reflected by different domains due to the partial coherence of the incident SR beam. Following the consideration in [53] we can write

$$C_{coh} = \exp[-\frac{1}{2}(D_{domains}/\xi_t)^2], \qquad (34)$$

where $\xi_t$ is the transverse coherence length of SR beam, $D_{domains}$ is the characteristic lateral domain size. If $D_{domains} \gg \xi_t$, $C_{coh} \approx 0$, the interference between scattering waves from different domains is absent. If $D_{domains} \ll \xi_t$, $C_{coh} \approx 1$ and scattering from the whole surface is fully coherent.

When $\vec{B}_{hf}$ orientates in the surface plane, it can be shown that in the kinematical approximation (remember that we consider M1 transition, so we have $\pi$ polarized SR beam with respect to the magnetic field of radiation)

$$f_\pi^L = f_\pi^{-L} = f_\pi^T = f_\pi^{-T} = f_\pi(t) \qquad (35)$$

$$f_\sigma^L = -f_\sigma^{-L} = f_\sigma(t); \quad f_\sigma^T = f_\sigma^{-T} = 0 \qquad (36)$$

So in the fully coherent scattering ($C_{coh} = 1$)

$$I_R(t) \propto \left((\alpha+\beta+\gamma+\delta)^2 |f_\pi(t)|^2 + (\alpha-\beta)^2 |f_\sigma(t)|^2\right), \qquad (37)$$

and when $\alpha = \beta = \gamma = \delta = 1$ no contribution of $|f_\sigma(t)|^2$ will be inserted to the time spectrum of reflectivity $I_R(t) \propto |f_\pi(t)|^2$. In the case of completely incoherent scattering ($C_{coh} = 0$)

$$I_R(t) \propto \left( (\alpha^2 + \beta^2 + \gamma^2 + \delta^2)|f_\pi(t)|^2 + (\alpha^2 + \beta^2)|f_\sigma(t)|^2 \right), \tag{38}$$

and when $\alpha = \beta = \gamma = \delta = 1$ we have $I_R(t) = A(|f_\pi(t)|^2 + \frac{1}{2}|f_\sigma(t)|^2)$. In general for any value of $C_{coh}$ we have

$$I_R(t) \propto (1 + 3C_{coh})|f_\pi(t)|^2 + \frac{1}{2}(1 - C_{coh})|f_\sigma(t)|^2 \tag{39}$$

The same mixture of $|f_\pi(t)|^2$ and $|f_\sigma(t)|^2$ dependencies we can get in the case of the homogeneous magnetization direction with definite azimuth angle $\gamma^{eff}$ for $\vec{B}_{hf}$ orientation:

$$I_R(t) \propto |f_\pi(t)|^2 + \sin^2 \gamma^{eff} |f_\sigma(t)|^2. \tag{40}$$

Compare (39) with (40) we can claim that the interpretation of the time spectra of reflectivity, presented in many papers in terms of uniaxial anisotropy of the sample magnetization (single domain state) can in reality be masked by the multidomain state due to the restricted coherence length of radiation. For any effective azimuth angle $\gamma^{eff}$, obtained by the fit of the reflectivity time spectrum, measured at one sample orientation, we can give a fully adequate description with some value of the partial coherence parameter $C_{coh}$ according to the equation for $C_{coh}$

$$\frac{(1 - C_{coh})}{2(1 + 3C_{coh})} = \sin^2 \gamma^{eff}. \tag{41}$$

In many cases the choice between the two different descriptions of the data can be made with the sample rotation say on 90°. If for the new spectrum interpretation you should change in the model the azimuth angle also for 90°, you deal with the single-domain state with definite magnetization direction. If you should not change anything in the model, them it is a proof of the partial coherence of the scattered waves from different domains. In the case of the external field application we should not rotate the sample but should change the direction of the external field, however in practice such comparison is not trivial, because the sample magnetization is always depends on the pre-history and you can get an another magnetization state in the process of increasing or decreasing of the external field.

Strictly speaking the lateral inhomogeneity always leads to the diffuse scattering. The "magnetic" diffuse scattering (or small angle scattering in the case of forward scattering), brightly demonstrated for NRS in [54, 55], gives the more direct information about the lateral magnetic domain size. However, the theory of the diffuse scattering especially for the case of the spectral inhomogeneity is rather complicated (see e.g. [56, 57]) and needs the use of the correlation functions for the description of the lateral distribution of the inhomogeneities, while for specular reflectivity this factor is absent and just the partial transverse coherence of synchrotron beam is actual. Diffuse NRR experiments are rather difficult to do (in the allocated beamtime) and the resulting advantage would have been of limited use for determination of the depth-profiles of magnetization direction. In our work we use just the specular NRR reflectivity and present the simplest and clear description of the influence of the finite transverse coherence of the synchrotron beam on the NRR from multi-domain multilayers. Note that similar simplification for description of the specular reflectivity from rough surfaces, like a simple exponential factor, decreasing the Fresnel amplitude of specular reflectivity, is commonly used for the X-ray reflectivity data treatment.

It is interesting that the surface domain influence on the specular NRR reflectivity destroys the common opinion that the specular reflectivity provides information on the depth profiles only, while off-specular scattering on the lateral structure of scattering layers. The same idea has been presented in [57], where the authors have shown that the lateral domain occurrence leads to the decrease of the "antiferromagnetic maximum" on the delayed specular NRR curve. However, in the NRR theory the authors of [57] have used a rather general parameter: "a specific magnetic bias parameter $\eta$" (eq. (45)), - which is a part of the magnetization along the field direction. In addition it has been assumed a strict antiferromagnetic interlayer coupling. So their approch would be helpless in the description of the twisted magnetization profiles.

**Experiment**

The experiment was performed at the station BL09XU of SPring-8. The angular dependencies and the time spectra of the NRR were measured at each step of the gradually increasing external field from 0 Oe up to 1500 Oe in L- and T- geometries. For such measurements the special design of the magnet had been created.

The [$^{57}$Fe (2.0 nm)/Cr (1.2 nm)]$_{10}$, [$^{57}$Fe(2.0 nm/Cr(1.2nm)]$_{20}$ and [$^{57}$Fe (3.0 nm)/Cr (1.2 nm)]$_{10}$ have been grown samples for our investigation at Indore (India). Deposition on Si substrate was carried out using ion beam sputtering at room temperature in a UHV chamber with a base pressure of 1 x 10$^{-7}$ mbar. Sputtering was done using 3 cm broad Kaufman type ion source with 1keV Ar ions. Fe layers were prepared with 95% enriched $^{57}$Fe target in order to enhance NRR signal. Before the measurements at Spring-8 the conversion electron Mössbauer spectra (CEMS) were measured (Fig. 3) as well as the magnetization curves (Fig. 4). CEMS can be fitted with several magnetic sub-spectra (we manage to do that with three sub-spectra used for the subsequent fit of the time spectra – Fig. 3). The ratio of the hyperfine lines confirms the plane anisotropy of the magnetization, i.e. all $\vec{B}_{hf}$ lay in the surface plane.

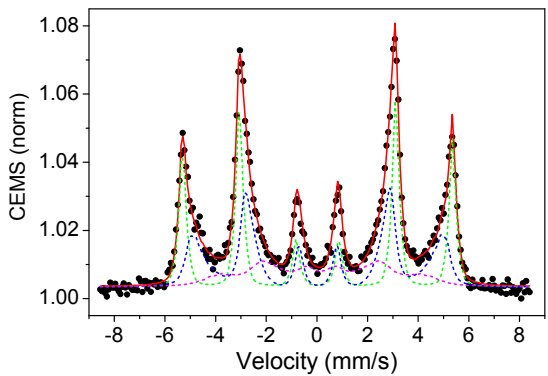

*Fig. 3. Experimental CEMS (dots) and fit result by three magnetic sub-spectra (dot lines) corresponding to $\vec{B}_{hf}^{(1)} = 33.1\,T$, $\vec{B}_{hf}^{(2)} = 30.8\,T$, $\vec{B}_{hf}^{(3)} = 24.5\,T$ for the [$^{57}$Fe (3.0 nm)/Cr (1.2 nm)]$_{10}$ sample.*

Magneto-optical Kerr effect (MOKE) measurements were done in longitudinal geometry using a He–Ne laser (632.8 nm wavelength), and with a maximum magnetic field of 1500 Oe.

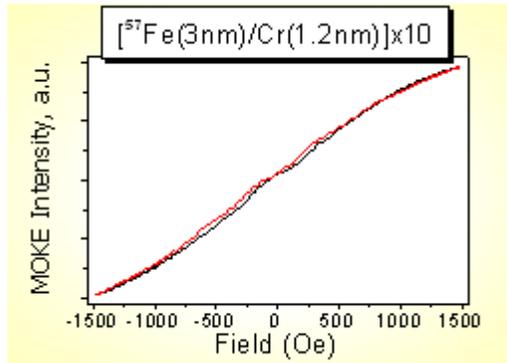

*Fig. 4. Magnetization curve for [$^{57}$Fe (3.0 nm)/Cr (1.2 nm)]$_{10}$ sample, measured by Magneto-Optical Kerr Effect.*

Magnetization measurements did not reveal the noticeable anisotropy in the surface plane. Just slight manifestation of the step-like behavior and weak hysteresis can be seen on the magnetization curves. The small values of these effects could be explained by the multi-domain surface structure of our film.

The antiferromagnetic interlayer coupling of $^{57}$Fe layers was evidently detected by the occurrence of the half order Bragg peaks (Fig. 5) on the delayed reflectivity curves measured at the SPring-8. We start these measurements with the sample [$^{57}$Fe(2.0 nm/Cr(1.2nm)]$_{20}$ and follow the disappearance of these "magnetic peaks" during the external field increase in the longitudinal geometry. However, we needed also the time spectra of reflectivity, but at the field decrease we come to another magnetization state of the sample. So later we decided to investigate the whole set of data in the increasing field with another [$^{57}$Fe (3.0 nm)/Cr (1.2 nm)]$_{10}$ sample. For that sample we got the most complete set of data, which we present now.

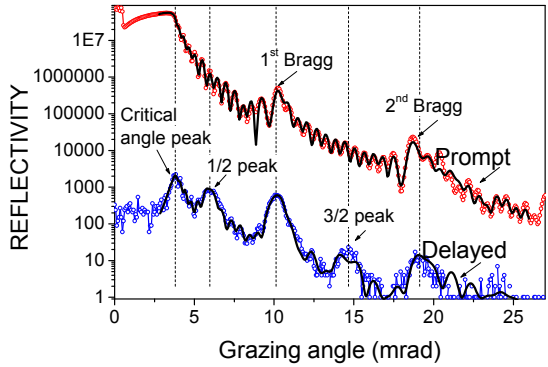

*Fig. 5. Prompt and delayed reflectivity measured without external field for the sample [$^{57}$Fe (3.0 nm)/Cr (1.2 nm)]$_{10}$. The angles at which the time spectra of the NRR were measured are marked by the dash vertical lines.*

The fit of the prompt reflectivity curve gives us the electron density profile (Fig. 6). It is not simple and we see that not only top Cr layer but also the surface Fe/Cr bilayer and bilayer at the interface with Cr buffer layer are somehow distracted. The electron density of Fe and Cr layers differs from the Table values due to some intermixture. The larger than table values of $\mathrm{Im}\,\chi$ in the interfaces can be explained by the roughness initiated diminution of radiation.

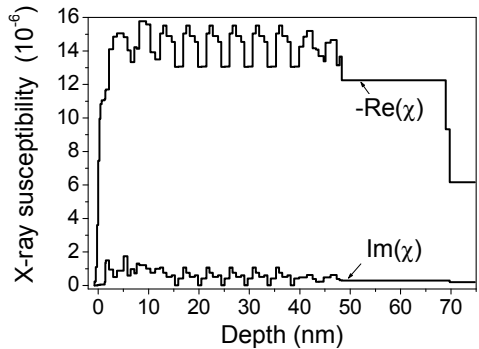

*Fig. 6. Depth profile of the electron density and of the absorption obtained by the fit of the prompt reflectivity curves.*

The fit of the delayed reflectivity curve should be done simultaneously with the fit of the time spectra of reflectivity which were measured at 5 angles: at the critical angle, at ½, 1$^{st}$.3/2 and 2$^{nd}$ orders Bragg peaks (Fig. 7). The shape of the time spectra measured at the marked grazing angles are rather different, however their Fourier transforms contain the same frequency beats and confirm that we have the plane orientation of $\vec{B}_{hf}$ when just the $\Delta m = \pm 1$ transitions are excited by the $\sigma$-polarized SR. Corresponding 4 lines in the conventional Mossbauer spectrum are shown in the insert of Fig. 7. The basic frequency beats originates from the interference of these hyperfine transitions (marked by vertical lines). At the same time the Fourier transform reveals the difference in the magnitude of these frequency beats. It have been shown that the first order Bragg peak is mostly sensitive to the hyperfine fields in the center of the $^{57}$Fe layers, but in 2$^{nd}$-order Bragg peak the contribution from interfaces dominates [23]. So the essential increase of the amplitude of the largest frequency beat for 2$^{nd}$-order Bragg peak can be explained by considerable disorientation of the $B_{hf}$ in the interfaces comparing with the center part of $^{57}$Fe layers. The same deduction have been obtained by the fit of the NRR data for [$^{57}$Fe/Cr] ML [43] and XRMR, PNR data for [Fe(35 Å)/Gd(50 Å)]$_5$ ML in [58].

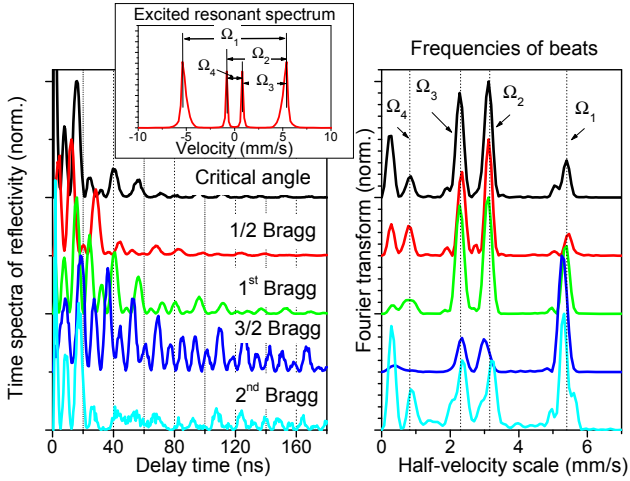

*Fig. 7. NRR time spectra, measured at the angles, marked by vertical lines in Fig. 5, (left panel) and their Fourier transform (right panel). The spectra are normalized and vertically shifted.*

For the joint fit of the delayed reflectivity and the time spectra of reflectivity we use the parameters, obtained by the Mössbauer spectrum fit and prompt reflectivity curve fit. The main purpose of the joint fit is the depth distribution of the three chosen $\vec{B}_{hf}$ and their effective orientation in plane.

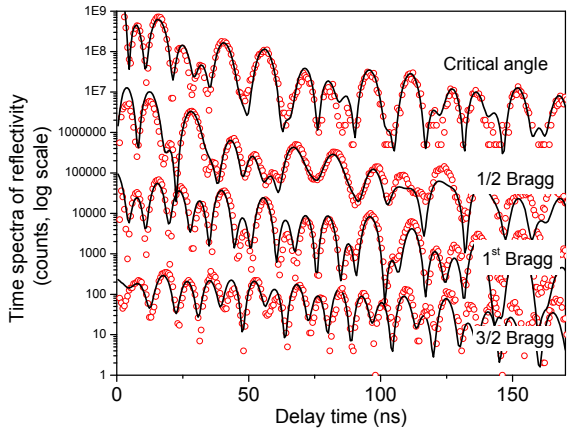

Fig. 8. Time spectra of NRR measured at 4 grazing angles in logarithm scale and vertically shifted. Symbols present the experimental data, lines – the fit results.

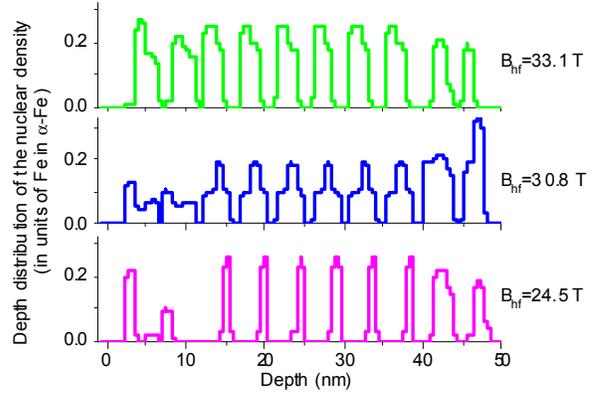

Fig. 9. The depth distribution of the $^{57}Fe$ nuclei, characterizing by one of the three kinds of $B_{hf}$, obtained by the fit of CEM spectrum in Fig. 3.

The fitted delayed curve and the time spectra of reflectivity in the case, when the external field is absent, are shown in Figs. 5,8. The obtianed depth-profiles for the three chosen hyperfine fileds is presented in Fig. 9.

As we see in Fig. 9 the highest hyperfine field of 33.1 T is attributed to the nuclei situated preferably near the top interface. That means that the Fe-on-Cr interface is more diffused than the Cr-on Fe interface in our sample.

We expected that when the external field is absent the hyperfine fields, antiferromagnetically coupled between adjacent $^{57}Fe$ layers, has no any preferable azimuth direction in the surface plane. Even if there is some uniaxial anisotropy in the multilayer due to internal stresses generated during deposition (like it was observed in [59] for finemet ferromagnetic alloy) we can not exclude 180° lateral domains. The coherent averaging of scattering amplitudes in that case should give the effective azimuth angle 0°/180° for the magnetization direction in antiferromagnetic bilayers. However, the fit of our data set has been more or less successful with the azimuth angles of 20°/-160° (or equivalently  -20°/160°). The considered in the previous section influence of the partial coherence of the scattering from different domains can explain this result. Using (41) for the effective azimuth angle $\gamma^{eff} = 20\deg$, we get $C_{coh} = 0.45$, or $D_{domains}/\xi_t = 1.26$. The value of the transverse coherence length for ESRF source has been measured as ~ 3 μm in [52] (notice that it depends of the slit sizes), at grazing angle ~ 10 mrad the lateral coherence length is $\xi_t$ ~ 300 μm, so $D_{domains}$ ~ 400 μm looks as a quite true result.

**The results of the external field influence**

The results of the external field influence is presented by Figs. 10-15. In L-geometry the relative intensity of the half-order Bragg peak comparing with first-order Bragg peak is decreased with the increase of the field magnitude - Fig. 10. It could be interpreted by the gradual change of the antiferromagnetic interlayer alignment to the canted state as if the difference of the azimuth angles in the adjacent $^{57}$Fe layers changes from 180º at zero field to 0º at 1500 Oe. However the measurements in T-geometry are not consistent with such interpretation: at 600 Oe we have the essential increase of the half-order Bragg peak. Such effect has been observed earlier [55] and it was interpreted by the reorientation of the anitiferromagnetically coupled magnetizations to the perpendicular direction relative the external field – that is the bulk spin-flop transition. The same results have been presented in the interpretation of the PNR experiments (e.g. in [6,7]). For the case of the perpendicular orientation of the antiferromagnetically coupled ML we do not have any difference in the scattering amplitudes from the adjacent $^{57}$Fe layers in the L-geometry, but they have opposite sign in the T-geometry. So at half order Bragg peak we have zero sum for L-geometry and maximal sum for T-geometry when these waves from magnetic period are added with the $\pi$ space phase shift.

Quantitative interpretation of the layer magnetization reorientation by the action of the applied field has been obtained by the joint fit of the delayed reflectivity curve (Figs. 11,12) and three time spectra (Figs. 12,13) measured for each value of the applied field. The final picture of the magnetization alignment for all used magnitudes of the applied field is given in Figs. 14-15.

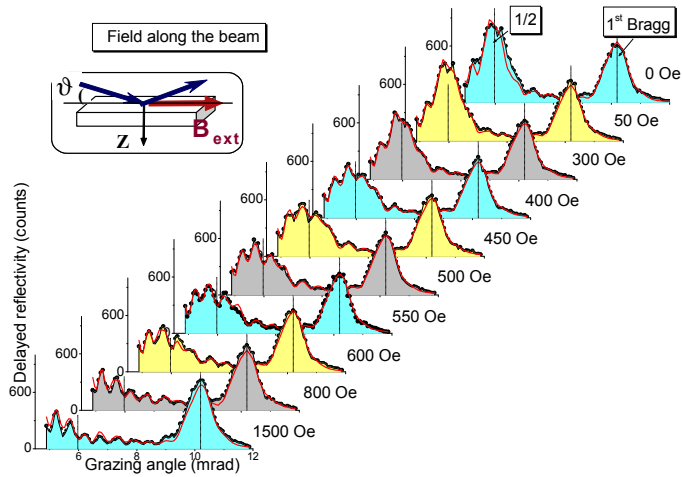
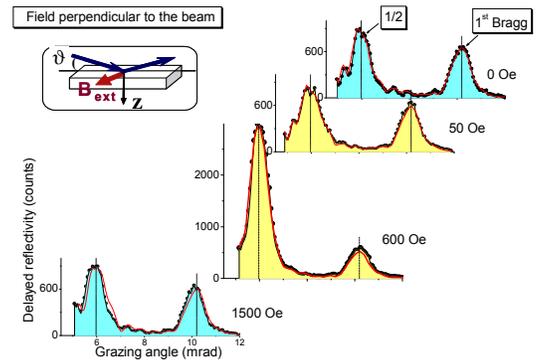

*Fig. 10. Delayed reflectivity curves measured at different magnitude of the ascending external field in L-geometry. Symbols represent the experimental data, lines – the fit results.*

*Fig. 11. The same as in Fig. 10 but for T-geometry.*

At first stage we supposed that in each magnetic sublattices (i.e. in the odd and even $^{57}$Fe layers) collinear alignment in each of two magnetic sublevels takes place. This approximate result is presented in Fig. 14 by dash vertical lines. It is significant to notice that in such approach we can get the more or less reasonable fit of the delayed reflectivity curves for all field magnitudes, but the obtained models does not reproduce all features of the experimental time spectra. For example, in Fig. 16 we compare the results for the half-order time spectra, measured under the applied 450 Oe field, obtained as the best joint fit of all the data for two models: the model of the collinear magnetization in each magnetic sublattice (dash vertical lines in Fig. 14) and for the model, allowing the arbitrary magnetization direction in each $^{57}$Fe layer. We see that the last more complicated model gives much better fit result. So we have used that model for all data.

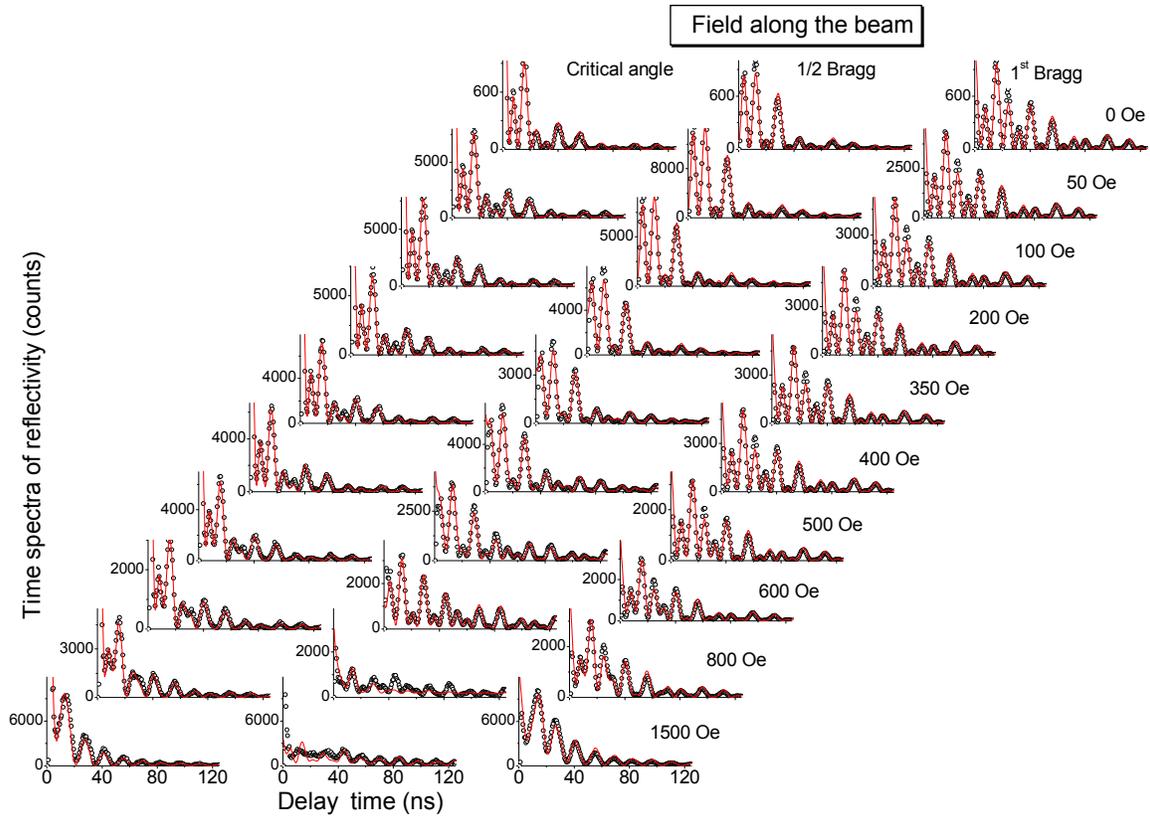

*Fig. 12. Time spectra of NRR reflectivity, measured at three incidence angles of SR for different magnitude of the ascending external field in L-geometry. Symbols represent the experimental data, lines – the fit results.*

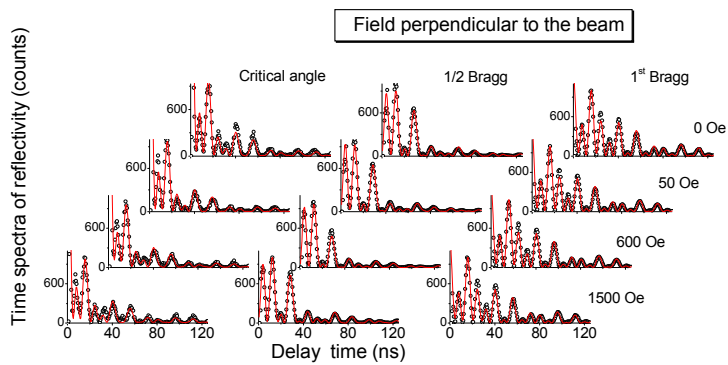

*Fig. 13. The same as in Fig. 12, but for T-geometry.*

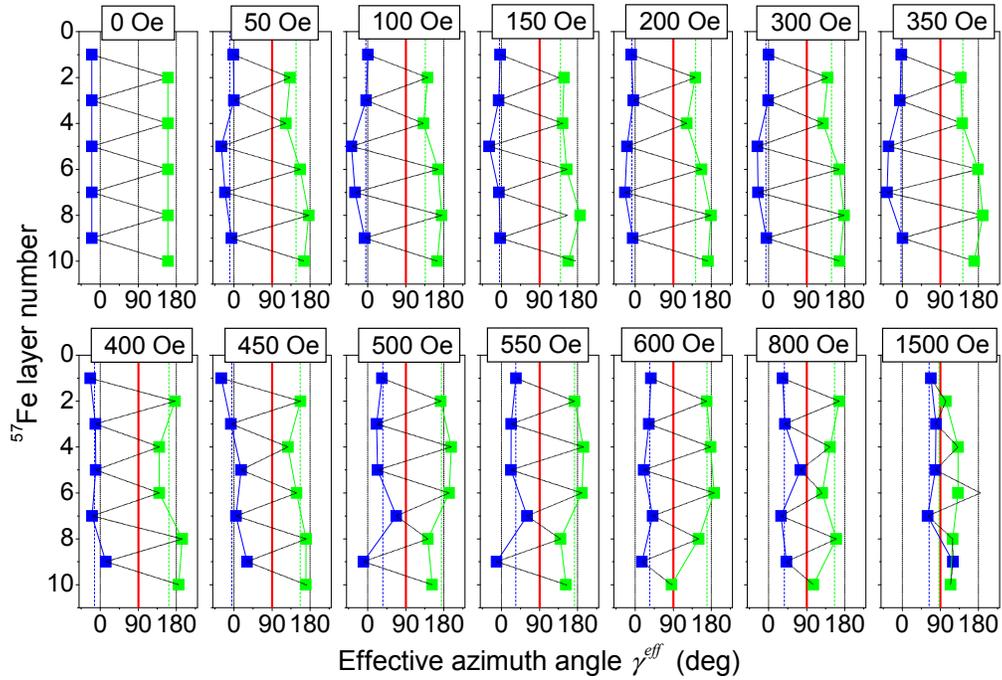

*Fig. 14. The layer-by layer variations of the effective azimuth angle in our ML as a function of the ascending applied field which are the results of the joint fit of the delayed reflectivity curves (Figs. 10-11) and time spectra of NRR reflectivity (Figs. 12-13).*

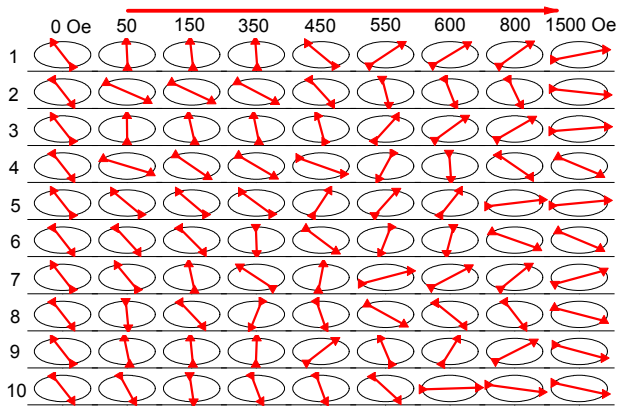

*Fig. 15. The same as in Fig. 14 (e.g. the magnetization directions in azimuth plane for 10 $^{57}$Fe layers of our ML) for the selected values of the external field, presented as polar graphs.*

Figs. 14-15 present the very complicated picture of the change of the layer-by-layer change of the effective azimuth angle for magnetization directions in $^{57}$Fe layers under the applied field. The most essential question now is how we can separate these effective angles onto the effect of the partially coherent averaging over magnetic domains and the real picture of the magnetization directions.

That can be done by the comparison of the fit result, obtained in the L- and T-geometries. Unfortunately the measurements in T-geometry have been done only at the selected field values (at 0, 50, 600 and 1500 Oe), and with the anther piece of the same sample in order to avoid the remanent magnetization effects after the first cycle of the magnetic field application (Figs. 11,13). The angular delayed curve and time spectra of reflectivity at zero external field are good fitted with the same hyperfine parameter and $B_{hf}$ orientations (20°/-160°) which according to (41) we associate with the $C_{coh} = 0.45$.

At 50 Oe the fit results for two geometries are slightly different (Fig. 17). It is almost the same in the simplest model of the collinear magnetizations in each magnetic sublattices (-9.6°/147° – the effective angles prove to be connected with the SR direction, but not with the applied field direction), so it is the $C_{coh}$ factor influence. The improved model show the twisted layer-by-layer magnetizations which are closer to the external field for near surface $^{57}$Fe layers than move off that direction and again slightly approach that direction at the bottom layers. And the marked twisted features are the same in both geometries, so this twisted part is connected with real change of the magnetization directions. However the influence of the multi-domain state is still predominant at 50 Oe, so the variations of the $\gamma^{eff}$ represent actually the average result over this multi-domain state.

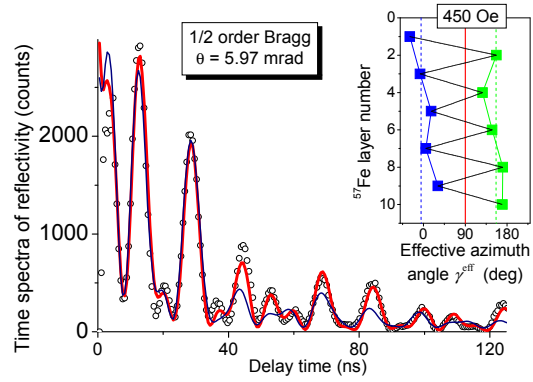

*Fig. 16. Time spectrum, measured at the ½ order Bragg peak from our sample with 450 Oe field applied along the beam (symbols) and theoretical curves for two models (being results of the best fit of all the data for this field): for the model of collinear magnetization in two "magnetic sublattices" -5º/156.7º- (thin blue line) and the model of twisted magnetization (thick red line), presented in the insert (thin dash vertical lines and lines with square symbols respectively).*

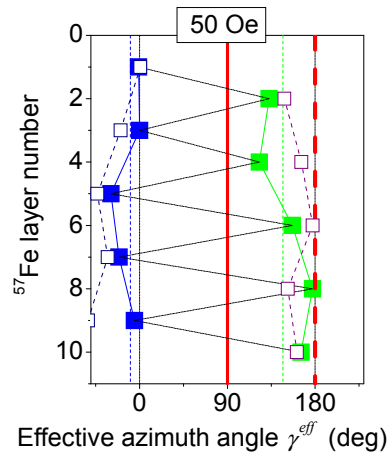

*Fig. 17. The layer-by layer variations of the effective azimuth angle in our ML for the applied field of 50 Oe obtained from the fit of data for L-geometry (filled aquares) and T-geometry (empty squares). The direction of the external field drawn by thick vertical line for the L-geometry and by thick dash vertical line for the T-geometry. Note that the effective azimuth angle $\gamma^{eff}$ is determined in the axis, connected with beam direction (Fig. 1).*

Contrary to the case of 50 Oe, the fit result for the 600 Oe and 1500 Oe in both geometries are almost identical: we should change the values of $\gamma^{eff}$, obtained by the fit of the data in L-geometry for the $\gamma^{eff} + 90^o$ (because $\gamma^{eff}$ is determined in the coordinate system, connected with beam direction) and the calculations in T-geometry give the acceptable result as well. That means that at these magnitudes of the external field (the bottom row in Fig. 14) we most probably have got the single domain state, and the obtained $\gamma^{eff}$ presents the real layer-by-layer magnetization directions.

At 600 Oe in T-geometry we have observed the very high half-order Bragg peak and in qualitative consideration we have supposed that in such a way the 'bulk" spin-flop effect manifests itself. The qualitative treatment of the data shows a noncollinear canted state. However even such ~ $140^o$ difference in magnetization directions for two magnetic sublattices (excuding bottom layers) is enough to create the high half-order Bragg peak contrary to the $(\gamma_1^{eff} + 90^o)/(\gamma_2^{eff} + 90^o) = 110^o/70^o$ at zero external field, because in the latter case it is not the real magnetization directions but the consequence of the partially coherent average of the scattering by different domains. We get the broadening of the angle between the magnetizations in the sublattices with external field increase getting the maximum at 350 Oe. That can be the evidence that the magnetizations in two sublattices really try to line up perpendicular to the external field at this field. But we have no the delayed reflectivity curve in T-geometry at such field. So we are not sure that the data at this magnitude of the field are completely stipulated by the real magnetization directions but not by the partial coherence of the beam.

At 1500 Oe the half-order Bragg peak is still presented in the T-geometry. So at the highest applied field we still have no total ferromagnetic alignment (as it could be concluded from the disappearance of the half-order Bragg peak in the L-geometry). It is interesting that the relative intensities of the half-order Bragg peak and the first order Bragg peak are almost the same at 0 Oe and at 1500 Oe. But the interpretation of these two dependencies is completely different. We believe that for 1500 Oe we get the real picture of the magnetization directions in Fig. 14, but at 0 Oe the specific azimuth angle of the antiferromagnetically coupled layers is the appearance of the partially coherent averaging of scattering from different magnetic domains.

The fit of the NRR data for our sample includes a huge amount of parameters (number of hyperfine parameters multiplied by the number of layers excluding the repetitions for the magnetization directions), so the problem of ambiguity of the presented picture in Fig. 14 is very serious. During the fit we have got sometimes the different models of magnetizations - Fig. 18. At first sight the obtained dependences presented by dash (blue on-line) and solid (red on-line) lines gives two completely different models. But soon we have understood that actually they are the same. They differ just by the sign of the external field, but our experimental results are not sensitive to the sign of the magnetization (the way for overcoming of this problem is described in e.g. [44]). So the observed two possible models just confirm that our fit is more or less reliable.

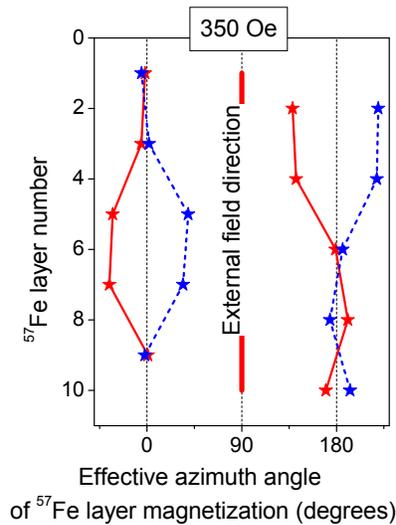

*Fig. 18. Two models of the depth profile of the $^{57}$Fe layer magnetizations under the 350 Oe applied field obtained by the all data joint fit.*

We have not analyzed the inhomogeneous intralayer magnetic structure, mentioned hereinabove. Probably such consideration can slightly improve the fit results, but it essentially increases the number of the fit parameters and make the task unsolvable. Besides just the second order Bragg peak is most sensitive to the interfaces (and their magnetization directions), but we have treated just the half- and first order Bragg peaks for different values of the applied field.

**Conclusions.**

In our experiment we have mobilized huge amount of the experimental data for the decryption of the magnetization reversal in antiferromagnetic [Fe/Cr] ML under the applied external field. The essential difference of our investigation from the previous ones (e.g. [6, 11, 16, 17]) consists in the involvement to the consideration the spectra of reflectivity (NRR time spectra in our case) measured at different Bragg angles in addition to the reflectivity curves. It is essential as well that the fit of all kinds of dependences has been done simultaneously for one and the same model.

For the first time we take into account the influence on the data interpretation the partial coherence of the scattering from different domains. We have shown that the fit of the reflectivity data can be performed only with the so-called "effective azimuth angle". The real meaning of this angle can be obtained just by comparison of the results obtained for two directions of the applied field relative the beam direction.

The obtained results, presenting in Figs.14-15, gives a rather complicated picture of the layer-by-layer resolved reorientation of magnetization in $^{57}$Fe layers under the applied field. The detailed analysis has shown that the collinear alignment in each magnetic sublattices and its cophasing rotation does not take place. We have seen that the reorientations even at the smallest applied field affected all layers but not just the top or bottom ones. The most specific magnetization state under the applied field is the twisted one, the bending details being the function of the applied field magnitude. The result should have some impact on the developing of the theory of the interlayer aniferromagnetic interaction. From our picture it is clear that in the theory we can not restrict ourselves by the interaction between just the adjacent magnetic layers, but should include the whole system simultaneously.

## ACKNOWLEDGMENTS

The work had been supported by RFBR grants No. 09-02-01293-a, 13-02-00760-a and 15-02-01502-a. Partial support through SERB grant SB/S2/CMP-007/2013 is thankfully acknowledged. The synchrotron radiation experiments were performed at SPring-8 with the approval of the Japan Synchrotron Radiation Research Institute (JASRI) (Proposal No. 2010B1298). The authors are very grateful to the Spring-8 administration for hospitality and perfect conditions for the experiment realization.

**References**

[1]    M. N. Baibich, J. M. Broto, A. Fert, F. Nguyen Van Dau, F. Petroff, P. Etienne, G. Creuzet, A. Friederich, and J. Chazelas, Phys. Rev. Lett. **61**, 2472 (1988).

[2]    http://www.nobelprize.org/nobel_prizes/physics/laureates/2007/

[3]    V. V. Ustinov, M. A. Milayev, L. N. Romashev, T. P. Krinitsina, A. M. Burkhanov, V. V. Lauter-Pasyuk, and H. J. Lauter, JMMM **300**, e281 (2006).

[4]    R. W. Wang, D. L. Mills, E. E. Fullerton, J. E. Mattson, and S. D. Bader, Phys. Rev. Lett. **72**, 920 (1994).

[5]    J. W. Freeland, V. Chakarian, Y. U. Idzerda, S. Doherty, J. G. Zhu, H. Wende, and C. C. Kao, Journal of vacuum science & technology. A. Vacuum, surfaces, and films, **16**(3), 1355 (1998).

[6]    V. Lauter-Pasyuk, H. J. Lauter, B. P. Toperverg, L. Romashev, and V. Ustinov, Phys. Rev. Lett. **89**, 167203-1-4 (2002).

[7]    S. G. E. te Velthuis, J. S. Jiang, S. D. Bader and G. P. Felcher, Phys. Rev. Lett. **89**, 127203 (2002).

[8]    S. Rakhmanova, D. L. Mills, and E. E. Fullerton, Phys. Rev. B **57**, 476 (1998).

[9]    L. Trallori, Phys. Rev. B **57**, 5923 (1998).

[10]    N. Papanicolaou, J. Phys. Condens. Matter **10**, L131 (1998); **11**, 59 (1999).

[11]    J. Meersschaut, C. L'abbé, F. M. Almeida, J. S. Ji-ang, J. Pearson, U. Welp, M. Gierlings, H. Maletta and S. D. Bader, Phys. Rev. B **73**, 144428-1-7 (2006).

[12]    V. V. Ustinov, N. G. Bebenin, L. N. Romashev, V. I. Minin, M. A. Milyaev, A. R. Del, and A. V. Semerikov, Phys. Rev. B **54**, 15958 (1996).

[13]    A. I. Morosov, A. S. Sigov, Uspekhi Fizicheskikh Nauk **180** (7), 709 (2010).

[14]    V.V. Ustinov, http://www.imp.uran.ru/UserFiles/File/dostizhenia/Ustinov.pdf

[15]    A. Nefedov, J. Grabis, and H. Zabel, Physica B **357**, 22 (2005).

[16]    T. Diederich, S. Couet, R. Röhlsberger, Phys. Rev. B **76**, 054401-1-5 (2007).

[17]	S. Couet,	K. Schlage,	Th. Diederich,	R Rüffer,	K. Theis-Bröhl, B. P. Toperverg, K. Zhernenkov, H. Zabel, and R. Röhlsberger, New Journal of Physics **11,** 013038 (2009).

[18]	G. V. Smirnov, U. van Bürck, , A. I. Chumakov, A. Q. R. Baron, and R. Rüffer, Phys. Rev. B **55**, 5811 (1997).

[19]	T. Mitsui, N. Hirao, Y. Ohishi, R. Masuda, Y. Nakamura, H. Enoki, K. Sakaki and M. Seto, J. Synchrotron Rad. **16**, 723 (2009).

[20]	V. Potapkin, A. I. Chumakov, G. V. Smirnov, J.-Ph. Celse, R. Rüffer, C. McCammon, and L. Dubrovinsky, J. Synchrotron Rad. **19**, 559 (2012).

[21]	G. V. Smirnov, Hyperfine Interactions **123/124,** 31 (1999).

[22]	A.I. Chumakov, L. Niesen, D.L. Nagy, and E.E. Alp, Hyperfine Interactions **123/124,** 427 (1999).

[23]	M.A. Andreeva and B. Lindgren, Phys. Rev. B **72**, 125422 (2005).

[24]	M. A. Andreeva, N. G. Monina, S. Stankov, Moscow University Physics Bulletin **63**(2), 132 (2008).

[25]	http://www.esrf.eu/computing/scientific/REFTIM/MAIN.htm

[26]	M.A. Andreeva, Hyperfine Interactions **185**, 17 (2008).

[27]	R. Azzam and N. Bashara, Ellipsometry and Polarized Light, North-Holland P.C. , 1977.

[28]	G. N. Borzdov, L. M. Barkovskii, and V. I. Lavrukovich, Zhurnal Prikladnoi Spectroskopii **25**, 526 (1976).

[29]	M.A. Andreeva, and K. Rosete, Vestnik Moscovskogo Universiteta, Fizika **41**(3), 65 (1986) (English transl. by Allerton press, Inc.); Poverkhnost', № 9, 145 (1986).

[30]	M. A. Andreeva, Journal of Physics: Conference Series **217**, 012013 (2010).

[31]	A. Angot, Complements de mathematiques, A l'usage des ingenieurs de l'elektrotechniques et des telecommunications, Paris 1957.§ 4.1.35.

[32]	N. K. Pleshanov, Z. Phys. B: Condens. Matter **94**, 233 (1994).

[33]	A. Ruhm, B. P. Toperverg, and H. Dosch, Phys. Rev. B **60**, 16073 (1999).

[34]	M. Blum,O. C. Kistner, Phys. Rev. **171**, 417 (1968).

[35]	I.W. Hamley, J.S. Pedersen, J. Appl. Cryst. **27**, 29 (1994).

[36]   M. A. Andreeva, Yu. L. Repchenko, Crystallography Reports **58**(7), 1037 (2013).

[37]   V. Holý, J. Kuběna, I. Ohlídal, K. Lischka, and W. Plotz, Phys.Rev. B **47**, 15896 (1993).

[38]   R. Röhlsberger, Hyperfine Interactions **123/124**, 301 (1999).

[39]   B. L. Henke, E. M. Gullikson, and J.C. Davis, Atomic Data and Nuclear Data Tables. **54** (2), 181 (1993).

[40]   G. N. Wertheim, Mossbauer effect. Principles and applications, Academic Press, New York and London, 1964.

[41]   E. Matthias, W. Schneider, and R.M. Steffen, Arkiv für Fysik **24**(9), 97 (1962).

[42]   D. L. Nagy, L. Bottyán, L. Deák, B. Degroote, J. Dekoster, O. Leupold, M. Major, J. Meersschaut, R. Rüffer, E. Szilágyi, A. Vantomme, Hyperfine Interactions **141/142**, 459 (2002).

[43]   Andreeva M.A., Semenov V.G., Häggström L., Kalska B., Lindgren B., Chumakov A.I., Leupold O. and Rüffer R., Hyperfine interactions **136/137**, 687 (2001).

[44]   C. L'abbe´ and J. Meersschaut, W. Sturhahn, J. S. Jiang, T. S. Toellner, E. E. Alp, and S. D. Bader, Phys. Rev. Lett. **93**, 037201 (2004).

[45]   T. Ślęzak, M. Ślęzak, M. Zając, K. Freindl, A. Kozioł-Rachwał, K. Matlak, N. Spiridis, D. Wilgocka-Ślęzak, E. Partyka-Jankowska, M. Rennhofer, A. I. Chumakov, S. Stankov, R. Rüffer, and J. Korecki, Phys. Rev. Lett. **105**, 027206 (2010).

[46]   J. Stöhr, H.A. Padmore, S. Anders, T. Stammler, M.R.Scheinfein, Surf. Rev. Lett. **5**  1297 (1998).

[47]   F. Nolting, A. Scholl, J. Stöhr, J. W. Seo, J. Fompeyrine, H. Siegwart, J.-P. Locquet, S. Anders, J. Lüning, E. E. Fullerton, M. F. Toney, M. R. Scheinfeink, and H. A. Padmore, Nature **405**, 767 (2000).

[48]   W. Kuch, Xingyu Gao, and J. Kirschner Phys. Rev. B **65**, 064406 (2002).

[49]   R. Röhlsberger, J. Bansmann, V. Senz, K.L. Jonas, A. Bettac, K.H. Meiwes-Broer, and O. Leupold, Phys. Rev. B **67**, 245412 (2003).

[50]    R. Röhlsberger, Nuclear Condensed Matter Physics with Synchrotron Radiation, Basic Principles, Methodology, and Applications (Springer, Berlin, 2004), Springer Tracts Mod. Phys., Vol. 208.

[51]    M.A. Andreeva, N.G. Monina, B. Lindgren, L. Häggström and B. Kalska, JETP **104** (4), 577 (2007).

[52]    A. Q. L. Baron, A. I. Chumakov, H. F. Grünsteudel, H. Grünsteudel, L. Niesen, and R. Rüffer, PRL **77**, 4808 (1996).

[53]    A. Q. L. Baron, Hyperfine Interactions **123/124**, 667 (1999).

[54]    Yu. V. Shvyd'ko, A. I. Chumakov, A. Q. L. Baron, E. Gerdau, R. Rüffer, A. Bernhard, and J. Metge, PRB **54**, 14942 (1996).

[55]    L. Bottyán, , J. Dekoster, L. Deák, B. Degroote, E. Kunnen, C. L'abbé, G.Langoucher, O. Leupold, M. Major, J. Meersschaut, D. L. Nagy, R. Rüffer, ESRF Highlights – 1999, p.62.

[56]    R. M. Osgood III, S. K. Sinha, J. W. Freeland, Y. U. Idzerda, and S. D. Bader, JMMM **198-199**, 698 (1999).

[57]    L. Deák, L. Bottyán, D. L. Nagy, H. Spiering, Yu. N. Khaidukov, and Y. Yoda, PRB 76, 224420 (2007).

[58]    E. Kravtsov, D. Haskel, S. G. E. te Velthuis, J. S. Jiang, and B. J. Kirby, Phys. Rev. B **79**, 1334438 (2009).

[59]    P. Sharma, A. Gupta, Nuclear Instruments and Methods in Physics Research B **244**, 105 (2006).

Figure captions

Fig. 1. The used coordinate system.

Fig. 2. The illustration of the partially coherent interference of waves reflected by different domains.

Fig. 3. Experimental CEMS (dots) and fit result by three magnetic sub-spectra (dot lines) corresponding to $\vec{B}_{hf}^{(1)} = 33.1\,\text{T}$, $\vec{B}_{hf}^{(2)} = 30.8\,\text{T}$, $\vec{B}_{hf}^{(3)} = 24.5\,\text{T}$ for the [$^{57}$Fe (3.0 nm)/Cr (1.2 nm)]$_{10}$ sample.

Fig. 4. Magnetization curve for [$^{57}$Fe (3.0 nm)/Cr (1.2 nm)]$_{10}$ sample, measured by Magneto-Optical Kerr Effect.

Fig. 5. Prompt and delayed reflectivity measured without external field for the sample [$^{57}$Fe (3.0 nm)/Cr (1.2 nm)]$_{10}$. The angles at which the time spectra of the NRR were measured are marked by the dash vertical lines.

Fig. 6. Depth profile of the electron density and of the absorption obtained by the fit of the prompt reflectivity curves.

Fig. 7. NRR time spectra, measured at the angles, marked by vertical lines in Fig. 5, (left panel) and their Fourier transform (right panel). The spectra are normalized and vertically shifted.

Fig. 8. Time spectra of NRR measured at 4 grazing angles in logarithm scale and vertically shifted. Symbols present the experimental data, lines – the fit results.

Fig. 9. The depth distribution of the $^{57}$Fe nuclei, characterizing by one of the three kinds of B$_{hf}$, obtained by the fit of CEM spectrum in Fig. 3.

Fig. 10. Delayed reflectivity curves measured at different magnitude of the ascending external field in L-geometry. Symbols represent the experimental data, lines – the fit results.

Fig. 11. The same as in Fig. 10 but for T-geometry.

Fig. 12. Time spectra of NRR reflectivity, measured at three incidence angles of SR for different magnitude of the ascending external field in L-geometry. Symbols represent the experimental data, lines – the fit results.

Fig. 13. The same as in Fig. 12, but for T-geometry.

Fig. 14. The layer-by layer variations of the effective azimuth angle in our ML as a function of the ascending applied field which are the results of the joint fit of the delayed reflectivity curves (Figs. 10-11) and time spectra of NRR reflectivity (Figs. 12-13).

Fig. 15. The same as in Fig. 14 (e.g. the magnetization directions in azimuth plane for 10 $^{57}$Fe layers of our ML) for the selected values of the external field, presented as polar graphs.

Fig. 16. Time spectrum, measured at the ½ order Bragg peak from our sample with 450 Oe field applied along the beam (symbols) and theoretical curves for two models (being results of the best fit of all the data for this field): for the model of collinear magnetization in two "magnetic sublattices" -5°/156.7°- (thin blue line) and the model of twisted magnetization (thick red line), presented in the insert (thin dash vertical lines and lines with square symbols respectively).

Fig. 17. The layer-by layer variations of the effective azimuth angle in our ML for the applied field of 50 Oe obtained from the fit of data for L-geometry (filled aquares) and T-geometry (empty squares). The direction of the external field drawn by thick vertical line for the L-geometry and by thick dash vertical line for the T-geometry. Note that the effective azimuth angle $\gamma^{eff}$ is determined in the axis, connected with beam direction (Fig. 1).

Fig. 18. Two models of the depth profile of the $^{57}$Fe layer magnetizations under the 350 Oe applied field obtained by the all data joint fit.